\documentclass[journal]{IEEEtran}
\IEEEoverridecommandlockouts
\ifCLASSOPTIONcompsoc
 \usepackage[nocompress]{cite}
\elseArchitecture And Application Of Intelligent Logistics Systems For Steel Companies Based On IIoM Technology
 \usepackage{cite}
\fi
\ifCLASSINFOpdf
  \usepackage[pdftex]{graphicx}
  \usepackage{subfigure}
  \graphicspath{{figures/}{authors/}}
  \DeclareGraphicsExtensions{.pdf,.jpeg,.png}
\else
  \usepackage[dvips]{graphicx}
  \graphicspath{{../eps/}}
  \DeclareGraphicsExtensions{.eps}
\fi

\usepackage{booktabs}
\usepackage{url}
\usepackage{amsmath,amssymb,amsthm}
\usepackage{xcolor,bm,multirow,txfonts,stfloats}
\usepackage{epstopdf}
\usepackage{algorithm}
\usepackage{array}
\usepackage{algpseudocode}

\def\BibTeX{{\rm B\kern-.05em{\sc i\kern-.025em b}\kern-.08em
  T\kern-.1667em\lower.7ex\hbox{E}\kern-.125emX}}

\allowdisplaybreaks[4]

\begin{document}

\title{Hybrid stability augmentation control of multi-rotor UAV in confined space based on adaptive backstepping control}

\author{
\IEEEauthorblockN{Zhan Quanxi\IEEEauthorrefmark{1}},
\IEEEauthorblockN{Zhang Junrui\IEEEauthorrefmark{1}},
\IEEEauthorblockN{Sun ChenYang\IEEEauthorrefmark{1}},
\IEEEauthorblockN{Shen Runjie\IEEEauthorrefmark{1}},
\IEEEauthorblockN{He Bin\IEEEauthorrefmark{1}},\\
\IEEEauthorblockA{\IEEEauthorrefmark{1}Tongji University}
\thanks{Corresponding Author: Shen Runjie, E-mail:shenrunjie@tongji.edu.cn. }
}

\markboth{Hybrid stability augmentation control of multi-rotor UAV in confined space  based on adaptive backstepping control}%
{Hybrid stability augmentation control of multi-rotor UAV in confined space  based on adaptive backstepping control}

\maketitle

\begin{abstract}
This paper applies the UAV to the inspection of water diversion pipelines in hydropower stations. The diversion pipeline is an enclosed space, so the airflow disturbance caused by the rotation of the UAV blades and the strong air convection from the chimney effect have a great impact on the flight control of the UAV. Although the traditional linear control PID flight control algorithm has been widely used and can meet the requirements of general flight tasks, it cannot guarantee the stability of the system over a wide range. The inspection of a diversion line in an enclosed space requires high system stability and robustness of the UAV controller. In this paper, a hybrid stabilised adaptive backstepping control method is proposed. Firstly, a multi-rotor UAV model is analysed and transformed into a strict feedback form with external disturbances; then adaptive techniques are used to estimate the airflow disturbances caused by the blades, and the attitude and position tracking controllers are designed by combining backstepping control and PID control respectively; finally, the asymptotic stability of the system is ensured by constructing a Lyapunov function. The experimental data show that the flight controller designed in this paper has good robustness and tracking performance, and can better resist the disturbance caused by airflow disturbance in confined space.
\end{abstract}



%
\IEEEpeerreviewmaketitle

\section{Introduction}
\label{sec1}
In recent years, compared with fixed-wing UAVs, multi-rotor UAVs have unique flight capabilities such as vertical takeoff and landing and hovering in the air. It has been widely used in agriculture, industry, military, civil civil applications such as search  and rescue~\cite{1,2}, disaster response~\cite{3,4}, safety monitoring~\cite{5,6}, infrastructure inspection~\cite{7,8}, precision agriculture~\cite{9,10}, exploration and mapping~\cite{11,12}.  The rapid development of multi-rotor UAVs has played an irreplaceable role in these fields.  A multi-rotor UAV is a non-linear system with non-linear, statically unstable, underdynamic, inter-axis coupling, multiple input and multiple output characteristics~\cite{13}. A quadrotor UAV has six degrees of freedom when flying in space (displacement motion in three directions in a rectangular coordinate system and rotational motion about three axes of the coordinate system). However, there are only four controllable variable inputs, making the control complex and susceptible to uncertainty. The design and implementation of high quality flight controllers has been a new challenge due to the tedious parameter tuning of non-linear control and the performance analysis involved.

Since the development of multi-rotor aircraft,there are three main types of flight control algorithms: 1)  Flight control method based on linear flight control theory~\cite{14}; 2) Learning-based flight control method~\cite{15}; 3) Model-based nonlinear control method~\cite{16}. At present, the flight control of multi-rotor UAV systems mainly adopts linear controller or combined with an optimization algorithm to improve the fast response and steady-state error of control to improve flight control performance. The flight control methods based on linear flight control theory include the classical cascade PID method~\cite{18}. Classical cascade PID control algorithms are one of the most successful and widely used control methods, for example, classical PID control ~\cite{18}, fuzzy PID controller~\cite{19}, neural network based PID control ~\cite{20}. Pixhawk flight control~\cite{21} is a cascade PID controller and has achieved a good control effect,which is the most classical and general flight control method at present; The possible reasons are as follows: 1. After tuning, the cascade PID controller can meet the needs of conventional flight missions; 2. The principle of PID controller is intuitive, simple structure and easy to implement; Although the PID controller occupies a high proportion in practical application, it is a single input single output controller designed for hovering balance point, which can meet the requirements of general flight missions, but can not ensure the large-scale stability of the system. In the case of extensive external interference, the nonlinear characteristics of the controlled object will cause the decline of control quality, and the cascade PID controller generally includes multiple control loops. For the new model without setting experience, there are problems of cumbersome parameter setting and strong experience dependence.

With the rapid development of sensors and drone technology, humans can use multi-rotor drones to carry data collection equipment to inspect hard-to-reach areas. For example, the Swiss company flyability's Elios drone ~\cite{22} inspects sewers and performs maintenance on many indoor industrial facilities. A full inspection takes 20-30 minutes, but the drone's endurance is only 7 minutes. Vijay R. Kumar at the University of Pennsylvania studied the automated flight of drones in dam aqueducts over five years, from semi-autonomous in 2014 to fully automated in 2019~\cite{23,24,25,26}.However, the UAVs developed by the team had a short endurance and detected the lower bend and lower horizontal sections of the dam aqueduct, rather than the entire area of the diversion pipeline. In this paper, the entire area of the Three Gorges Hydropower Station diversion pipeline was inspected. The automatic inspection area includes:
\begin{itemize}
\item Lower horizontal section.
\item Lower bend section.
\item Oblique straight section.
\item Upper bend section.
\item Upper flat section.
\end{itemize}

The presence of strong air convection from the chimney effect and turbulent wind fields caused by the rotation of the UAV blades inside the hydropower station's diversion pipes can cause significant disruption to the UAV's flight control. A high quality controller with solid stability and robustness over a wide range needs to be designed for the UAV.As the GPS signal is shielded in the confined space of the diversion pipeline, the positioning of the UAV depends on sensors such as LIDAR, optical flow and IMU. The navigation method on UAV positioning is not the focus of this paper and is not described in detail in this paper. In this paper, a high quality controller based on a hybrid stabilisation method with adaptive backstepping control is proposed. First, a model of a quadrotor UAV is developed and transformed into a strict feedback form with uncertainty. Then, an adaptive method is used to estimate the upper limit of delay due to large airflow disturbances and an adaptive backstepping controller is designed for attitude loop control. a PID controller is used to control the altitude and position of the aircraft. Finally, the stability of the closed-loop system is analysed in conjunction with Lyapunov stability theory. A dual closed-loop controller with an inner-loop attitude and an outer-loop position achieves hybrid stability-enhanced control of a multi-rotor UAV.

\section{Basic Model of Multi-rotor UAV}
\label{sec2}
\begin{figure}[ht!]
\centering
\subfigure[The body coordinate system.]{\includegraphics[height=3cm]{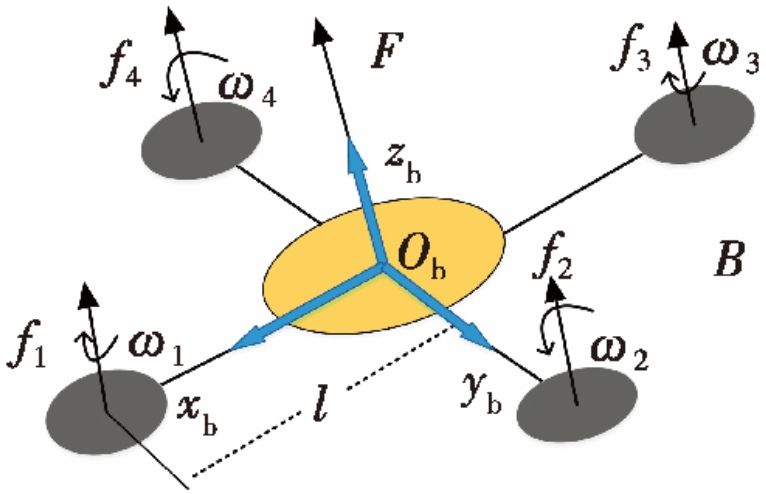}%
\label{fig1a}}\qquad
\subfigure[The geographic coordinate system.]{\quad\includegraphics[height=2.7cm]{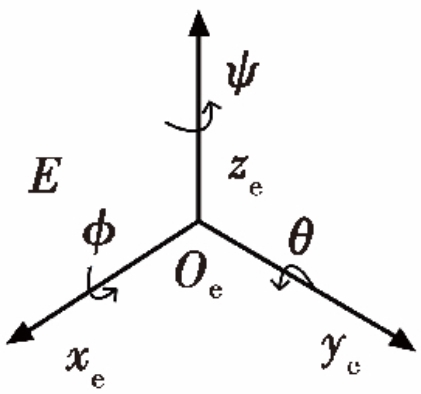}%
\label{fig1b}}
\caption{{The body coordinate system and the geographic coordinate system of the UAV.}\label{fig1}}
\end{figure}

The quadrotor UAV is the most typical multi-rotor UAV and the dynamics model of the quadrotor UAV is used as an example in this paper.The quadrotor  UAV can establish body coordinate system $\bm{B}(O_b-x_b y_b z_b)$ and geographic coordinate system $\bm{E}(O_e-x_e y_e z_e)$. In the body coordinate system, the coordinate origin $(O_b)$ is located in the center of mass of the drone. The axis $x_b$ and axis $y_b$ point to the center of the propellers 1 and 2 respectively, and the axis $(z_b)$ is oriented perpendicular to the plane upwards, as shown in Fig~\ref{fig1a}. In the geographic coordinate system, the coordinate origin $O_e$ is located at the takeoff point of the quadrotor UAV, and the positive direction of $(x_e,y_e,z_e)$ can be recorded as the north, east and up respectively, as shown in Fig~\ref{fig1b}. 

The position and attitude of the drone in the geographic coordinate system are $\bm{\xi}=[x,y,z]^T$ and $\bm{\eta}=[\phi,\theta,\psi]^T$, where the roll angle $(\phi)$, pitch angle $(\theta)$, and yaw angle $(\psi)$ of the drone are represented, respectively. In the body frame, the four-rotor UAV's line velocity  and angular velocity  are represented by $\bm{v}=[v_x,v_y,v_z]^T$ and $\bm{\varpi}=[\varpi_{xb},\varpi_{yb},\varpi_{zb}]^T$.

Then, the kinematic model of the quadrotor UAV can be represented as:
\begin{equation}
\label{eq1}
\left\{
\begin{array}{@{}l}
\dot{\bm{\xi}}=R_t\bm{v}\\
\dot{\bm{\eta}}=R_t\bm{\varpi}
\end{array}
\right.
\end{equation}

Among them, the line velocity transfer matrix $R_t$ indicates the rotation relationship of the body coordinate system relative to the geographical coordinate system, and the angular velocity transfer matrix $R_r$ represents the rotational relationship between the velocity vector and Euler angular velocity in the body coordinate system.The velocity vector is $\bm{\varpi}=[\varpi_{xb},\varpi_{yb},\varpi_{zb}]^T$ and the Euler angular velocity is $\dot{\bm{\eta}}=[\dot{\phi},\dot{\theta},\dot{\psi}]^T$. According to the literature~\cite{28,27}, the specific expressions $R_t$, $R_r$ are (2) and (3):
\begin{equation}
\label{eq2}
R_t=
\begin{bmatrix}
c\theta c\psi &s\phi s\theta c\psi-c\phi s\psi &c\phi s\theta c\psi+s\phi s\psi \\
c\theta s\psi &s\phi s\theta c\psi+c\phi s\psi &c\phi s\theta c\psi-s\phi s\psi \\
-s\theta &s\phi c\theta &c\phi c\theta \\
\end{bmatrix}
\end{equation}
\begin{equation}
\label{eq3}
R_r=
\begin{bmatrix}
1&0&-s\theta \\
0&c\phi &c\theta s\phi \\
0&-s\phi &c\theta c\phi
\end{bmatrix}
\end{equation}

Among them, $c\theta$ is $\cos\theta$, $c\psi$ is $\cos\psi$, $c\phi$ is $\cos\phi$, $s\theta$ is $\sin\theta$, $s\psi$ is $\sin\psi$, $s\phi$ is $\sin\phi$.

The kinematic model of the multi rotor UAV is composed of translational motion and rotation. As shown in formula (\ref{eq1}), it represents the motion characteristics of the position and attitude of the multi rotor UAV.  However, the kinematic equations only reflect changes in the state of motion of the UAV and do not include the force and moment factors that cause the changes in motion, such as strong air convection due to the chimney effect and turbulent wind field perturbations due to paddle rotation in the diversion pipeline.The flight controller of UAV needs a dynamic model to describe the motion including wind disturbance. Assuming that the drone is a rigid body and the center of mass is the origin of the body coordinate system, ignoring the gyro effect of the blade, the translation equation and rotation equation are finally established from Newtonian mechanics or Euler Lagrange equation, such as literature~\cite{27,28}.The dynamic model of four rotor UAV is shown in equations (\ref{eq4}) and (\ref{eq5}):
\begin{equation}
\label{eq4}
\left\{
\begin{array}{@{}l}
m\ddot{x}=u_1 (c\psi s\theta c\phi+s\psi s\phi)-K_f \dot{x}\\
m\ddot{y}=u_1 (s\psi s\theta c\phi-s\psi s\phi)-K_f \dot{y}\\
m\ddot{z}=u_1 c\theta c\phi-K_f \dot{z}-g
\end{array}
\right.
\end{equation}
\begin{equation}
\label{eq5}
\left\{
\begin{array}{@{}l}
I_x \ddot{\phi}=\dot{\theta}\dot{\psi}(I_y- I_z)+J_r \dot{\theta}\Omega_r +lu_2-d_{\phi}\\
I_y \ddot{\theta}=\dot{\theta}\dot{\phi}(I_z- I_x)-J_r \dot{\phi}\Omega_r +lu_3-d_{\theta}\\
I_z \ddot{\psi}=\dot{\theta}\dot{\phi}(I_x- I_y)  +lu_4-d_{\psi}
\end{array}
\right.
\end{equation}

Among them, l is the distance between the quadrotor center of mass and  the rotation axis of propeller, $m$ is the total load weight of the UAV; $I_x$, $I_y$, $I_z$ are the rotational inertia around each axis; $K_f$ is the wind disturbance coefficient;$J_r$ is the total rotational inertia of the entire motor rotor and propeller around the body's rotational axis;$d_i(i=/phi,/theta,/psi) is the airflow disturbance moment$; $u_1$ is the total thrust of the UAV, $u_2$ is the roll motion control torque, $u_3$ is the pitch motion control torque, $u_4$ is the yaw motion control torque.$\Omega_r$ is the combined speed of each rotor of the drone, it satisfy the following relationship:
\begin{equation}
\label{eq6}
\Omega_r=-\varpi_1-\varpi_3+\varpi_2+\varpi_4
\end{equation}

\begin{figure}[!ht]
\centering
\includegraphics[width=\columnwidth]{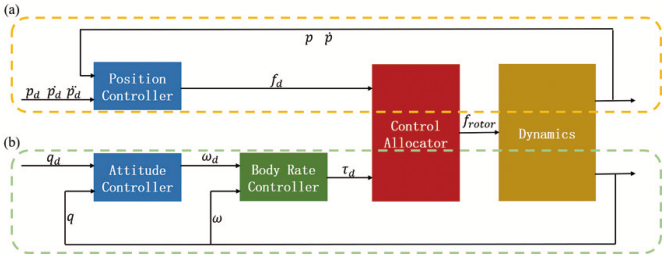}
\caption{Flight control algorithm structure}
\label{fig2}
\end{figure}

The flight control algorithm of the UAV can be divided into internal and external rings, including the attitude ring and position ring, as shown in Figure~\ref{fig2}.

Corresponding dividing the formula (\ref{eq4}) and (\ref{eq5}) into two-part subsystems
\begin{equation}
\label{eq7}
\left\{
\begin{array}{@{}l}
m\ddot{x}=u_1 (c\psi s\theta c\phi+s\psi s\phi)-K_f \dot{x}\\
m\ddot{y}=u_1 (s\psi s\theta c\phi-s\psi s\phi)-K_f \dot{y}
\end{array}
\right.
\end{equation}
\begin{equation}
\label{eq8}
\left\{
\begin{array}{@{}l}
m\ddot{z}=u_1 c\theta c\phi-K_f \dot{z}-g\\
I_x \ddot{\phi}=\dot{\theta}\dot{\psi}(I_y-I_z)+J_r \dot{\theta}\Omega_r +lu_2-d_{\phi}\\
I_y \ddot{\theta}=\dot{\psi}\dot{\phi}(I_z- I_x)-J_r \dot{\phi}\Omega_r +lu_3-d_{\theta}\\
I_z \ddot{\psi}= \dot{\theta}\dot{\phi}(I_x-I_y) +lu_4-d_{\psi}
\end{array}\right.
\end{equation}

From (\ref{eq7}) and (\ref{eq8}),it can be seen that the controller consists of two parts: the attitude controller and the position controller.When the total rotor lift u1 is a definite value, the displacement acceleration depends on the magnitude of the attitude angle, so the attitude angle can determine the flight path. Therefore, the general design idea of the control system is:
\begin{itemize}
\item The position controller receives the desired position($x_d,y_d,z_d$) and desired yaw($\psi_d$) from the input and outputs the desired attitude($\phi_d,\theta_d$) and total lift of the UAV($u_1$)
\item The attitude controller receives the input desired attitude and outputs the attitude control torque($u_2,u_3,u_4$) of the UAV.
\item The motor distribution model and the motor efficiency model convert the total lift control and the attitude control torque ($u_1,u_2,u_3,u_4$) of the UAV into the speed of each rotor of the UAV and thus control the motion of the UAV.
\end{itemize}
The objective of this paper is therefore to design a control law for the control  total lift control and the attitude control torque ($u_1,u_2,u_3,u_4$) of the UAV

\section{Design a hybrid stability-increasing controller based on adaptive backstepping}
\label{sec3}
\subsection{Attitude controller design}
\label{sec3.1}
The following assumptions are made before designing the controller:

\textbf{Assumption 1}: During the movement of UAVs, ($x_d$, $y_d$, $z_d $), ($\phi_d$, $\theta_d$, $\psi_d $) are continuously steerable.

\textbf{Assumption 2}: For the airflow disturbance moment uncertain term $d_i$ ($i=\phi,\theta,\psi$), there is a constant $r_i>0$, so that $|d_i| \leqslant r_i$

(1)Pitch angle subsystem design:

Firstly,the pitch angle equation of state model in the dynamics model equation (5) is transformed into a strict feedback form so that it meets the backstepping controller design requirements.
\begin{equation}
	\label{eq9}
	\left\{
	\begin{array}{@{}l}
		{x_1}=\theta\\
		{x_2}=\dot{\theta} 
	\end{array}
	\right.
\end{equation}
\begin{equation}
	\label{eq10}
	\left\{
	\begin{array}{@{}l}
		\dot{x_1}=x_2\\
		\dot{x_2}=a_1 U_3+f_{\theta}+d_{\theta}
	\end{array}
	\right.
\end{equation}

Among them, $a_1=\frac l{I_y}$, ${f}_{\theta} =\dot{\phi}\dot{\psi}\frac{(I_z- I_z)}{I_y}+\frac{J_r}{I_y}\dot{\phi}\Omega_r$

The tracking error of the roll angle is:
\begin{equation}
\label{eq11}
\mathrm{e}_1=\theta_d-\theta
\end{equation}

There is the derivative for the error $\mathrm{e}_1$:
\begin{equation}
\label{eq12}
\dot{\mathrm{e}_1}=\dot{\theta_d}-\dot{\theta}
\end{equation}

This paper constructs Lyapunov function$V_1$:
\begin{equation}
\label{eq13}
V_1=\frac12 \mathrm{e}_1^2
\end{equation}

There is the derivative for the $V_1$:
\begin{equation}
\label{eq14}
\dot{V_1}=\mathrm{e}_1 \dot{\mathrm{e}_1}=\mathrm{e}_1(\dot{\theta_d}-\dot{\theta}) 
\end{equation}

To ensure the stability of the system, i.e. $\dot{V_1} < 0$, introduce the virtual control quantity $x_v$.
\begin{equation}
 \label{eq15}
 x_v=c_1\mathrm{e}_1+\dot{\theta_d}
\end{equation}
where $c_1$ is a positive constant and there is an error between $x_v$ and $x_2$, noted as $\mathrm{e}_2$.
\begin{equation}
	\label{eq16}
	\mathrm{e}_2=x_2-x_{v}=-c_1\mathrm{e}_1-\dot{\mathrm{e}_1}
\end{equation}
To ensure the stability of the system $V_1$ at $\mathrm{e}_1$ = 0, this paper constructs another the Lyapunov function $V_2$:
\begin{equation}
	\label{eq17}
	V_2=V_1+\frac12 \mathrm{e}_2^2+\frac1{2\beta_1} \tilde{r_1}^2
\end{equation}

In the formula (\ref{eq17}), $\hat{r_1}$ is the estimation of $r_1$, $\tilde{r_1}$ is the estimation error of $r_1$($\tilde{r_1}=r_1-\hat{r_1}$), $\beta_1$ is a positive adaptive constant.

There is the derivative for the Lyapunov function $V_2$:
\begin{align}
\label{eq18}
\dot{V_2}&=\dot{V_1}+\mathrm{e}_2 \dot{\mathrm{e}_2}-\frac1{\beta_1} \tilde{r_1}\dot{\hat{r_1}}\notag\\
&=-c_1 \mathrm{e}_1^2+\mathrm{e}_2 (\dot{x_2}-\mathrm{e}_1-\dot{x_{v}})-\frac1{\beta_1} \tilde{r_1}\dot{\hat{r_1}}\notag\\
&=-c_1 \mathrm{e}_1^2+\mathrm{e}_2 (a_1 U_3+f_{\theta}+{d_{\theta}}-\mathrm{e}_1-\dot{x_{v}})-\frac1{\beta_1} \tilde{r_1}\dot{\hat{r_1}}
\end{align}
In order to ensure the stability of the pitch angle subsystem,i.e. $\dot{V_2} < 0$, the control law of the pitch angle controller is designed as formula (\ref{eq19}).
\begin{equation}
	\label{eq19}
	\left\{
	\begin{array}{@{}l}
		U_3=\frac1{a_1}(\mathrm{e}_1+\dot{x_{v}}-f_{\theta}-\hat{d_\theta}-c_2\mathrm{e}_2)\\
		\dot{\hat{d_\phi}}=\beta_1\mathrm{e}_2
	\end{array}
	\right.
\end{equation}
Where, $\beta_1>0$, $c_1>0$, $c_2>0$.
Substituting the above control law equation (19) into equation (17) yields:
\begin{equation}
	\label{eq20}
	\dot{V_2}= -c_1 \mathrm{e}_1^2-c_2 \mathrm{e}_2^2 \leqslant 0
\end{equation}
Therefore,from Lyapunov stability theory, the pitch angle controller subsystem designed in this paper is stable.

(3)Design of roll angle controller and yaw angle controller subsystems

Based on formulas (\ref{eq7}) and (\ref{eq8}), The Roll control torque $u_2$ and yaw control torque $u_4$ can be designed:
\begin{equation}
	\label{eq21}
	\left\{
	\begin{array}{@{}l}
		U_2=\frac1{a_2}(\mathrm{e}_3+\dot{x_{2v}}-f_{\phi}-\hat{d_\phi}-c_4\mathrm{e}_4)\\
		\dot{\hat{d_\phi}}=\beta_2\mathrm{e}_4
	\end{array}
	\right.
\end{equation}
\begin{equation}
	\label{eq22}
	\left\{
	\begin{array}{@{}l}
		U_4=\frac1{a_3}(\mathrm{e}_5+\dot{x_{3v}}-f_{\psi}-\hat{d_\psi}-c_6\mathrm{e}_6)\\
		\dot{\hat{d_\psi}}=\beta_3\mathrm{e}_6
	\end{array}
	\right.
\end{equation}

In the formula (\ref{eq21})and (\ref{eq22}) , the expected roll angle and yaw angle are $\phi_d$ and $\psi_d$. The tracking error is $\mathrm{e}_3=\phi_d-\phi$, $\mathrm{e}_5=\psi_d-\psi$, :
\begin{equation}
\label{eq23}
V_3=\frac12 \mathrm{e}_3^2+\frac12 \mathrm{e}_4^2+\frac1{2\beta_2}\tilde{r_2}^2,\quad V_4=\frac12 \mathrm{e}_5^2+\frac12 \mathrm{e}_6^2+\frac1{2\beta_3}\tilde{r_3}^2
\end{equation}

Refer to the analysis process of $V_2$ above, and we can get:
\begin{equation}
\label{eq24}
\dot{V_3}\leqslant -c_3 \mathrm{e}_3^2-c_4 \mathrm{e}_4^2 \leqslant 0, \quad \dot{V_4}\leqslant -c_5 \mathrm{e}_5^2-c_6 \mathrm{e}_6^2 \leqslant 0
\end{equation}

In the  formula (\ref{eq24}), $\mathrm{e}_4=-c_3\mathrm{e}_3-\dot{\mathrm{e}_3}$, $\mathrm{e}_6=-c_5\mathrm{e}_5-\dot{\mathrm{e}_5}$, $c_3$, $c_4$, $c_5$, $c_6>0$, so the attitude subsystem is asymptotically stable.

\subsection{Position PID controller design}
\label{sec3.2}
The stability controller designed in this paper uses a double closed-loop structure: PID control for the height and horizontal position of the UAV and the backstepping adaptive control for the attitude control of the UAV.

Taking $P=(x_d,y_d,z_d,\psi_d)$ input control parameters, three virtual control parameters  $(U_x,U_y,U_z)$ can be obtained:
\begin{equation}
\label{eq25}
U_x=K_{px} (x_d-x)+K_{dx} (x'_d-x')
\end{equation}
\begin{equation}
\label{eq26}
U_y=K_{py} (y_d-y)+K_{dy} (y'_d-y')
\end{equation}
\begin{equation}
\label{eq27}
U_z=K_{pz} (z_d-z)+K_{dz} (z'_d-z')
\end{equation}

To make full use of acceleration information and strengthen adjustment, this article adopts the following form:
\begin{equation}
\label{eq28}
U_x=K_{px} (x_d-x)+k_{dx} (x'_d-x') +K_{ddx} x''
\end{equation}
\begin{equation}
\label{eq29}
U_y=K_{py} (y_d-y)+k_{dy} (y'_d-y') +K_{ddx} y''
\end{equation}
\begin{equation}
\label{eq30}
U_z=K_{pz} (z_d-z)+k_{dz} (z'_d-z')+K_{ddx} z''
\end{equation}

Taking $(U_x, U_y, U_z)$ as the resultant force in the three-axis directions of the $E(X,Y,Z)$ system, from Newton's second law, the following formula can be obtained:
\begin{equation}
\label{eq31}
\begin{bmatrix}
0\\
0\\
mg
\end{bmatrix}
+R_t
\begin{bmatrix}
0\\
0\\
u_1
\end{bmatrix}
=m
\begin{bmatrix}
a_x\\
a_y\\
a_z
\end{bmatrix}
\end{equation}

From the above formula (\ref{eq31}):
\begin{equation}
\label{eq32}
\begin{bmatrix}
a_x\\
a_y\\
a_z
\end{bmatrix}
=\frac1m
\begin{bmatrix}
(c\phi s\theta c\psi+s\phi s\psi)u_1\\
(c\phi s\theta s\psi-s\phi c\psi)u_1\\
c\phi c\theta u_1-mg
\end{bmatrix}
\end{equation}

Then the three-axis resultant force is:
\begin{equation}
\label{eq33}
\begin{bmatrix}
U_x\\
U_y\\
U_z
\end{bmatrix}
=\frac1m
\begin{bmatrix}
(c\phi s\theta c\psi+s\phi s\psi)u_1\\
(c\phi s\theta s\psi-s\phi c\psi)u_1\\
c\phi c\theta u_1-mg
\end{bmatrix}
\end{equation}

For the UAV inspection process, the UAV always keeps the nose still, that is, $\psi_d$ is a fixed value, which can be inversely solved to obtain:
\begin{equation}
\label{eq34}
u_1=m\sqrt{U_x^2+U_y^2+(U_z+g)^2}
\end{equation}
\begin{equation}
\label{eq35}
\theta_d=\frac{\arcsin[U_x m-u_1 s\phi_d s\psi_d]}{u_1 c\psi_d c\phi_d}
\end{equation}
\begin{equation}
\label{eq36}
\phi_d=\arcsin[U_x s\psi_d-U_y c\psi_d] \frac{m}{u_1}
\end{equation}

In the process of controlling the pitch angle to move along the x-axis, it can be assumed that the pitch $(\psi)$ and yaw $(\phi)$ angles of the UAV are fixed at this time, so $a_x= (f_1\sin\theta-f_2)\frac{u_1}m$, where $f_1=c\phi_d c\psi_d$, $f_2=s\phi_d s\psi_d$, both are constants, then
\begin{align}
\label{eq37}
\theta&=\arcsin\bigg\{\frac{u_1}m\bigg[f_1\bigg[K_{px} (x_d-x)+K_{ix}\int(x_d-x) dt+K_{dx} \dot{x_d}\notag\\
&\quad~-\dot{x})\bigg]+f_2\bigg]\bigg\}
\end{align}

For the process of controlling the roll angle along the y-axis direction, a derivation similar to the above can be done, and this article will not repeat it.

For the height control of the UAV, the PID control is linear, and the tracking error is $\mathrm{e}_7=z_d-z$, where $z_d$ is the desired height, then:
\begin{equation}
\label{eq38}
a_z=\frac1m c\phi c\theta u_1-g
\end{equation}

Where $g$ represents the local acceleration of gravity, when the drone is hovering, the pitch and roll angles of the drone are 0, which can be simplified to:
\begin{equation}
\label{eq39}
a_z=\frac{u_1}m-g
\end{equation}

From the above formula, the height direction control input is:
\begin{equation}
\label{eq40}
u_1=[K_{pz} (z_d-z)+K_{iz} \int(z_d-z)\mathrm{d}t+K_{dz} (\dot{z_d}-\dot{z})]-mg
\end{equation}

According to the above attitude loop design functions (\ref{eq19}) (\ref{eq21}) (\ref{eq22}) and Lyapunov function proof of the stability , combined with the position and height loop design functions (\ref{eq35})(\ref{eq36})(\ref{eq40}), a high-quality UAV controller in a confined space is finally obtained.
\section{Experiment}
\label{sec4}
This paper uses a self-assembled multi-rotor UAV as the airborne platform. The flight control uses CuavV5+, and the attitude loop is rewritten based on the open-source APM firmware. The attitude loop controller adopts the adaptive backstepping control method described above. The drone's airborne platform is equipped with Velodyne16 lidar for environmental information perception. The Insta360-OneX2 camera collects image information for 3D pipeline reconstruction and defect recognition; The onboard processor of DJI Manifold 2-C is used for multi-sensor fusion and navigation, VIJIM-VL66 fill light enhances lighting, and other equipment, as shown in Figure~\ref{fig3} below. The water pipe diameter is 12.4~m, and the elevation is 64.3~m. The specific details of the model are shown in Figure~\ref{fig4} below. 
\begin{figure}[!ht]
\centering
\includegraphics[width=\columnwidth]{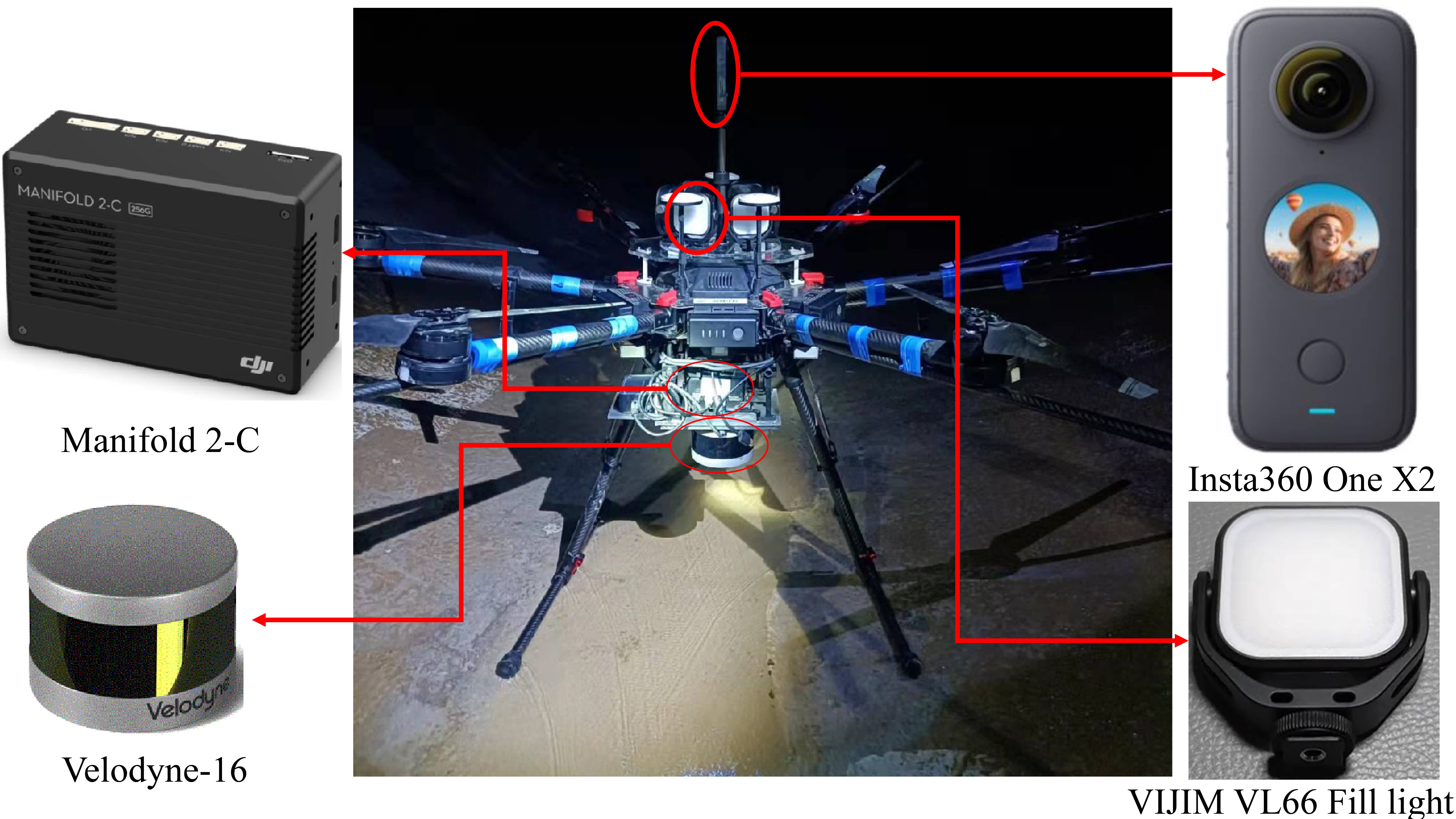}
\caption{Schematic diagram of drone equipment}
\label{fig3}
\end{figure}
\begin{figure}[!ht]
\centering
\includegraphics[width=\columnwidth]{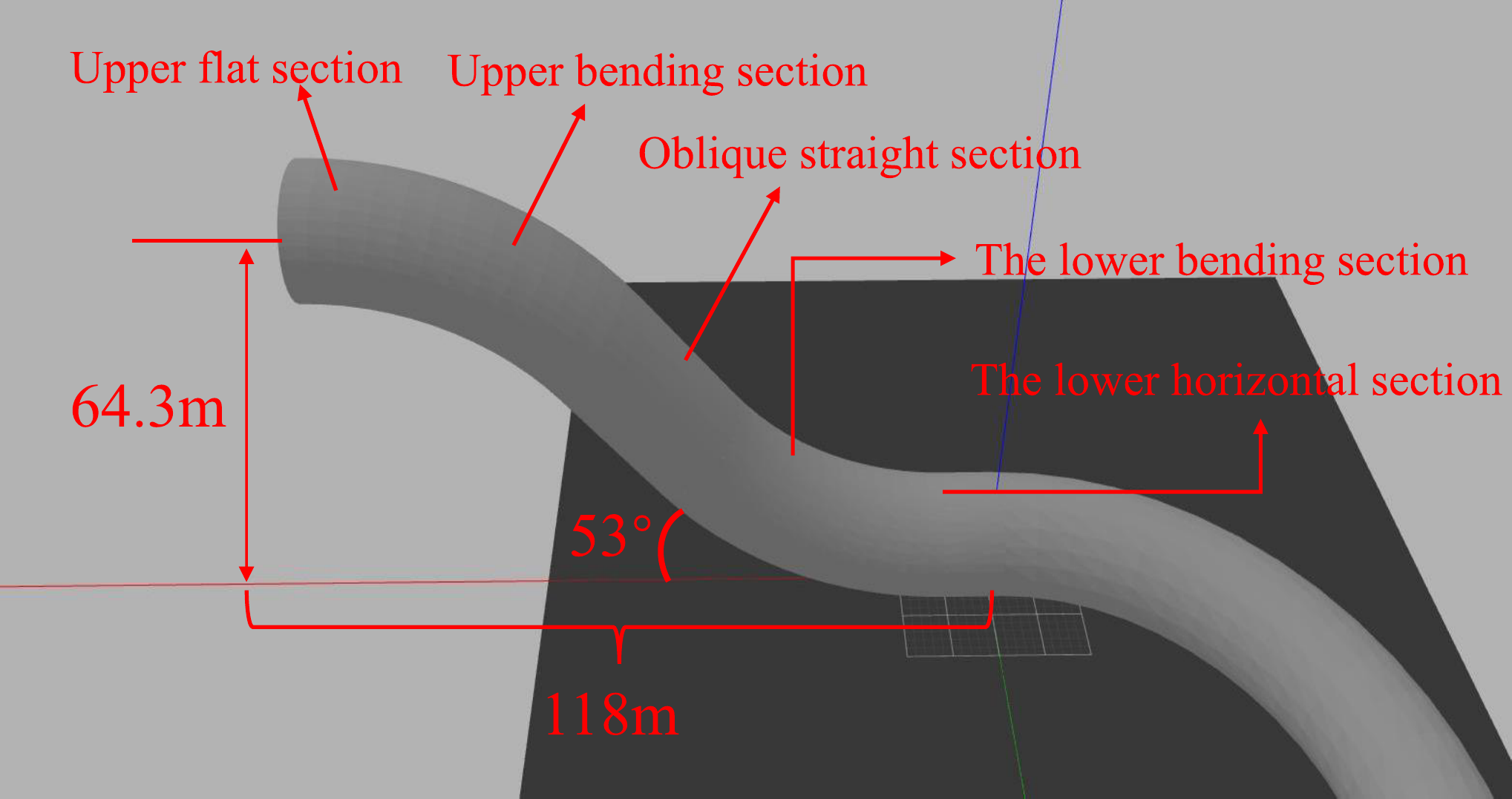}
\caption{Diversion pipeline model of hydropower station}
\label{fig4}
\end{figure}

The experimental scenario in this paper is a water diversion pipeline of a hydropower station, and the inside of the water diversion pipeline is GPS rejection space. The GPS position cannot be obtained as the true value to compare the control effects of the position loop controller. In order to obtain the actual reference value of the position loop controller, the test is carried out under the condition of outdoor wind force 5 (wind speed 8m$/$s-10m$/$s), and the expected position and actual position of the position ring are compared, as shown in Figure~\ref{fig5} below. This paper uses the following three parameters to evaluate the position loop control effect: 1. The average value of the tracking error of the position loop is the average value of the difference between the expected position and the actual position; 2. The standard deviation of the tracking error; 3. The tracking error The percentage of the mean value of the actual position relative to the mean value of the actual position. The position loop's x-direction and y-direction horizontal directions use the same parameters. This article uses the x-direction expected position and actual position to compare the control effect in the horizontal direction. The comparison diagram of the horizontal desired position and the actual position is shown in Figure~\ref{fig6}(a). The percentage of the tracking error relative to the average actual position is shown in Figure~\ref{fig6}(c). The comparison chart between the desired height position and the actual height position is shown in~\ref{fig6}(b) below, the percentage of the tracking error relative to the average actual position, As shown in Figure~\ref{fig6}(d).
\begin{figure}[!ht]
\centering
\includegraphics[width=\columnwidth]{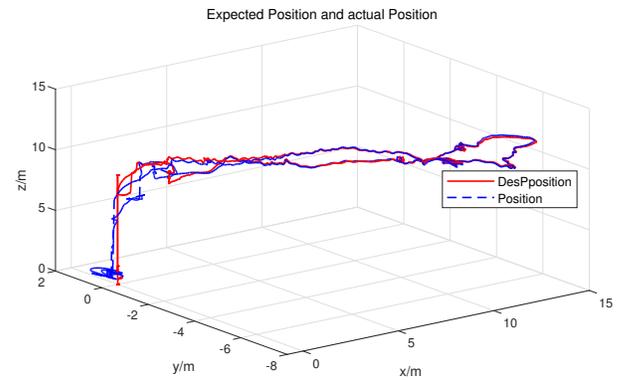}
\caption{Comparison of expected and actual positions of drones}
\label{fig5}
\end{figure} 
\begin{figure}[!ht]
\centering
\includegraphics[width=\columnwidth]{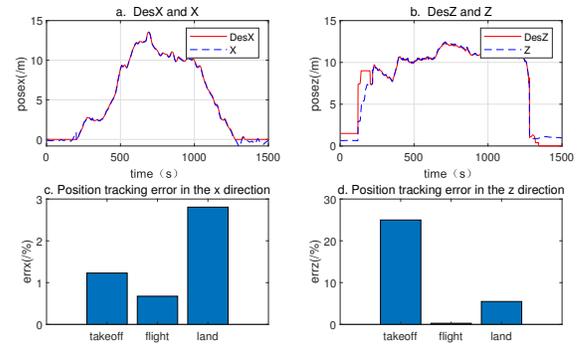}
\caption{Comparison of tracking errors in the horizontal direction and the height direction}
\label{fig6}
\end{figure} 
\begin{table}[!ht]
\centering
\caption{The evaluation index of the position loop control effect in the horizontal direction}
\label{tab1}
\tabcolsep=2pt
\begin{tabular}{cm{1.6cm}<{\centering}m{1.8cm}<{\centering}m{4cm}<{\centering}}
\toprule
&Mean tracking error(/m)&The standard deviation of tracking error&The percentage of the mean value of the tracking error relative to the mean value of the actual position\\
\midrule
Takeoff &0.1228&0.1225&1.3\%\\
Flight&0.0444&0.0872&0.47\%\\
Land&0.272&0.522&2.88\%\\
\bottomrule
\end{tabular}
\end{table}
\begin{table}[!ht]
\centering
\caption{Evaluation index of position loop control effect in the height direction}
\label{tab2}
\tabcolsep=2pt
\begin{tabular}{cm{1.6cm}<{\centering}m{1.8cm}<{\centering}m{4cm}<{\centering}}
\toprule
&Mean tracking error(/m)&The standard deviation of tracking error&The percentage of the mean value of the tracking error relative to the mean value of the actual position\\
\midrule
Takeoff&1.665&0.3218&26.67\%\\
Flight&0.204&0.1067&3.26\%\\
Land&0.337&0.2349&5.43\%\\
\bottomrule
\end{tabular}
\end{table}

This article divides the drone's flight into three stages: 1. 0--200~s is the takeoff phase; 2. 200--1250~s is the flight phase; 3. 1250--1500~s is the landing phase. The z-direction starts to be pulled up to the desired height while keeping the x-direction motionless; during the landing phase, the z-direction gradually reduces the desired height to 0 and keeps the x-direction still.As can be seen from figures~\ref{fig5} and~\ref{fig6}, even in the case of level 5 wind interference, the actual position of the drone and the desired position follow better in the flight phase, the drone achieves an ideal control effect. During the takeoff and landing phases, the drone will have large fluctuations that will likely cause the UAV to roll over, which is a relatively dangerous moment. From Tables~\ref{tab1} and~\ref{tab2}, it can be seen that the tracking error in the takeoff and landing phases is large, and the tracking error in the Z direction is bigger than that in the X-direction. For the x-direction, the desired position during the takeoff phase is always 0, and the drone's position during the unlocked takeoff will have a sudden change. The drone's position during the landing phase will be unstable and fluctuate to a certain extent, resulting in large errors. The UAV can fly smoothly during the flight phase. The mean value and standard deviation of the tracking error are small, providing a better guarantee for the UAV attitude loop control. For the Z direction, in the UAV before unlocking, its desired position is 2~m, and the actual position of the drone is about 0.5~m. After the drone is unlocked, its desired position becomes 8m, and then it is pulled up to 9~m, and the corresponding drone will climb upward, which gradually reaches the desired position. As a result, the difference between the desired position in the climbing phase and the actual position in the z-direction is large, and the average tracking error is enormous. In the flight phase, the UAV has better followability, with the three evaluation indicators of the attitude control loop are all small, and can achieve the desired control effect. The three evaluation indicators in the landing phase are smaller than those in the takeoff stage. The positioning error mainly causes the tracking error in the height direction of the UAV.

In this paper, the flight data inside the water pipeline of the Three Gorges Hydropower Station is used to evaluate the control effect of the attitude loop. When taking off, the wind speed at a distance of 2m around the UAV measured by the anemometer is about 6.5m/s.Three parameters similar to the evaluation of the position loop are used as evaluation indicators for the evaluation of the attitude loop control effect: 1. the mean value of the attitude loop's tracking error is the mean value of the difference between the expected attitude and the actual attitude; 2. the standard deviation of the attitude tracking error; 3. the percentage of the average attitude tracking error relative to the actual average attitude. As shown in Table~\ref{tab3} below, it is the evaluation index of the attitude loop control effect.
\begin{table}[!ht]
\centering
\caption{Evaluation index of the position loop control effect in the height direction}
\label{tab3}
\tabcolsep=2pt
\begin{tabular}{cm{1.9cm}<{\centering}m{2.3cm}<{\centering}m{3.5cm}<{\centering}}
\toprule
&Mean value of attitude tracking error (/deg)&The standard deviation of attitude tracking error&The percentage of the average tracking error relative to the average actual attitude\\
\midrule
pitch&0.028&0.3218&0.16\%\\
roll&0.037&0.1067&0.32\%\\
yaw&0.193&0.2349&0.62\%\\
\bottomrule
\end{tabular}
\end{table}
\begin{figure}[!ht]
\centering
\includegraphics[width=\columnwidth]{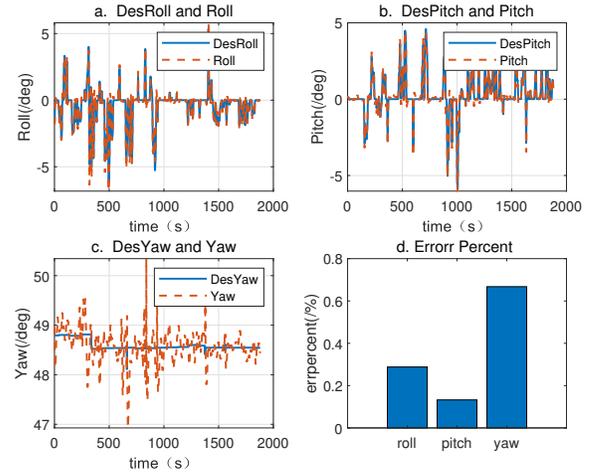}
\caption{Comparison of UAV attitude tracking error}
\label{fig7}
\end{figure}  

From the comparison of the expected attitude and the actual attitude in Figure~\ref{fig7} , the pitch angle and roll angle follow better, the mean value and standard deviation of the tracking error are small. The tracking error of the yaw angle is more extensive, and the convergence speed is relatively slow. This is because penstocks splice the water pipeline of the Three Gorges Hydropower Station, and there is a strong magnetic shield inside the pipeline. The magnetometer carried by the drone fails. This paper uses the yaw estimated by the RCFIC algorithm. The yaw has a specific error, so the yaw direction control's tracking error and standard deviation are larger than the pitch and roll angles. In order to avoid the tracking error fluctuation of the UAV's yaw from causing interference to the flight of the UAV, it is necessary to keep the UAV's yaw constant during the flight. From an overall point of view, the adaptive backstepping control adopted by the attitude loop controller has a better effect. Although there is a large airflow disturbance in the enclosed space of the water pipeline, the dynamic response is still good, and the overshoot and fluctuation of the actual position are slight. Realize the smooth flight of the UAV in the water diversion pipeline.

\section{Conclusion}
\label{sec5}
When the multi-rotor UAV is inspecting the water diversion pipeline of the hydropower station, because of the problem that the airflow caused by the rotation of the UAV blade has a greater impact on the flight motion, this paper proposes a hybrid stabilized flight algorithm with adaptive backstepping control. The algorithm designs the outer loop position controller into a PID controller, and the inner loop controller adopts adaptive backstepping control, which achieves hybrid stability enhancement by combining the two. The interior of the Three Gorges Hydropower Station is GPS-rejected space, and GPS positioning and navigation cannot be carried out, so the outer loop controller test was performed outdoors. The interior of the Three Gorges Hydropower Station is GPS-rejected space, and GPS positioning and navigation cannot be carried out, so the outer loop controller test was performed outdoors. Comparing the actual position of the GPS positioning and navigation system with the expected position, from the horizontal and altitude evaluation indicators in Table~\ref{tab1} and Table~\ref{tab2}, it can be seen that the real-time followability and error fluctuation of the UAV during the flight phase are slight. In contrast, during the takeoff and landing phases, the tracking error of the drone is relatively large, which is a dangerous phase in the drone's flight. In this paper, the attitude loop controller is tested in the water diversion pipeline of the Three Gorges Hydropower Station. The tracking attitude error of the pitch and roll angles is small. Due to the failure of the magnetic compass carried by the UAV, the yaw angle estimated by the RCFIC algorithm is used. There is a certain error, so the tracking attitude error and standard deviation of the yaw angle are relatively large. The water pipeline of the hydropower station is a confined space, the rotation of the blades of the UAV will cause large airflow disturbances during the flight. Still, the adaptive backstepping control hybrid stability control algorithm proposed in this paper can ensure that the UAV dynamic response is still good. The overshoot and fluctuation of the flight position are slight so that the UAV can fly smoothly in the water diversion pipe to reach an ideal robust and stable tracking control.



\bibliographystyle{IEEEtran}
\bibliography{mybib}

\begin{thebibliography}{10}
\providecommand{\url}[1]{#1}
\csname url@samestyle\endcsname
\providecommand{\newblock}{\relax}
\providecommand{\bibinfo}[2]{#2}
\providecommand{\BIBentrySTDinterwordspacing}{\spaceskip=0pt\relax}
\providecommand{\BIBentryALTinterwordstretchfactor}{4}
\providecommand{\BIBentryALTinterwordspacing}{\spaceskip=\fontdimen2\font plus
\BIBentryALTinterwordstretchfactor\fontdimen3\font minus
  \fontdimen4\font\relax}
\providecommand{\BIBforeignlanguage}[2]{{%
\expandafter\ifx\csname l@#1\endcsname\relax
\typeout{** WARNING: IEEEtran.bst: No hyphenation pattern has been}%
\typeout{** loaded for the language `#1'. Using the pattern for}%
\typeout{** the default language instead.}%
\else
\language=\csname l@#1\endcsname
\fi
#2}}
\providecommand{\BIBdecl}{\relax}
\BIBdecl

\bibitem{1}
E.~T. Alotaibi, S.~S. Alqefari, and A.~Koubaa, ``Lsar: Multi-uav collaboration
  for search and rescue missions,'' \emph{IEEE Access}, 2019.

\bibitem{2}
T.~Tomic, K.~Schmid, P.~Lutz, A.~Domel, M.~Kassecker, E.~Mair, I.~L. Grixa,
  F.~Ruess, M.~Suppa, and D.~Burschka, ``Toward a fully autonomous uav:
  Research platform for indoor and outdoor urban search and rescue,''
  \emph{IEEE robotics \& automation magazine}, 2012.

\bibitem{3}
N.~E. C. K. R. A. I.~F. Erdelj, M., ``Help from the sky: Leveraging uavs for
  disaster management,'' \emph{IEEE Pervasive Computing}, 2017.

\bibitem{4}
V.~Mayor, R.~Estepa, A.~Estepa, and G.~Madinabeitia, ``Deploying a reliable
  uav-aided communication service in disaster areas,'' \emph{Wireless
  Communications \& Mobile Computing}, 2019.

\bibitem{5}
N.~H. Motlagh, M.~Bagaa, and T.~Taleb, ``Uav-based iot platform: A crowd
  surveillance use case,'' \emph{Ieee Communications Magazine}, 2017.

\bibitem{6}
J.~L. Sun, F.~Liu, Y.~Z. Zhou, G.~Gui, T.~Ohtsuki, S.~Guo, and F.~Adachi,
  ``Surveillance plane aided air-ground integrated vehicular networks:
  Architectures, applications, and potential,'' \emph{Ieee Wireless
  Communications}, 2020.

\bibitem{7}
A.~K.~. Khattak~S, Papachristos~C, ``Change detection and object recognition
  using aerial robots,'' \emph{International Symposium on Visual Computing.},
  2016.

\bibitem{8}
Y.~P. Ma, Q.~W. Li, L.~L. Chu, Y.~Q. Zhou, and C.~Xu, ``Real-time detection and
  spatial localization of insulators for uav inspection based on binocular
  stereo vision,'' \emph{Remote Sensing}, 2021.

\bibitem{9}
V.~Puri, A.~Nayyar, and L.~Raja, ``Agriculture drones: A modern breakthrough in
  precision agriculture,'' \emph{Journal of Statistics \& Management Systems},
  2017.

\bibitem{10}
C.~Potena, R.~Khanna, J.~Nieto, R.~Siegwart, D.~Nardi, and A.~Pretto,
  ``Agricolmap: Aerial-ground collaborative 3d mapping for precision farming,''
  \emph{Ieee Robotics and Automation Letters}, 2019.

\bibitem{11}
T.~Dang, S.~Khattak, F.~Mascarich, K.~Alexis, and Ieee, ``Explore locally, plan
  globally: A path planning framework for autonomous robotic exploration in
  subterranean environments,'' in \emph{19th International Conference on
  Advanced Robotics (ICAR)}, 2019, Conference Proceedings.

\bibitem{12}
S.~Khattak, C.~Papachristos, and K.~Alexis, ``Vision-depth landmarks and
  inertial fusion for navigation in degraded visual environments,'' in
  \emph{13th International Symposium on Visual Computing (ISVC)}, 2018,
  Conference Proceedings.

\bibitem{13}
B.~Rao, A.~G. Gopi, and R.~Maione, ``The societal impact of commercial
  drones,'' \emph{Technology in Society}, 2016.

\bibitem{14}
M.~A. Elsadig and M.~Elbakri, ``Design of autopilot platform using hils
  approach,'' \emph{2017 International Conference on Communication, Control,
  Computing and Electronics Engineering (ICCCCEE)}, 2017.

\bibitem{15}
Kimathi, Kang’ethe, and Kihato, ``Application of reinforcement learning in
  heading control of a fixed wing uav using x-plane platform,''
  \emph{International Journal of Scientific \& Technology Research}, vol.~6,
  pp. 285--290, 2017.

\bibitem{16}
R.~Wang, L.~Gao, C.~R. Bai, and H.~Sun, ``U-model-based sliding mode controller
  design for quadrotor uav control systems,'' \emph{Mathematical Problems in
  Engineering}, 2020.

\bibitem{18}
Y.~Kartal, P.~Kolaric, V.~Lopez, A.~Dogan, and F.~Lewis, ``Backstepping
  approach for design of pid controller with guaranteed performance for
  micro-air uav,'' \emph{Control Theory and Technology}, 2020.

\bibitem{19}
J.~Dong and B.~He, ``Novel fuzzy pid-type iterative learning control for
  quadrotor uav,'' \emph{Sensors}, 2019.

\bibitem{20}
S.~Islam, P.~X. Liu, and A.~El~Saddik, ``Nonlinear adaptive control for
  quadrotor flying vehicle,'' \emph{Nonlinear Dynamics}, 2014.

\bibitem{21}
\BIBentryALTinterwordspacing
PIXHAWK. [Online]. Available: \url{http://www.Pixhawk.com}
\BIBentrySTDinterwordspacing

\bibitem{22}
\BIBentryALTinterwordspacing
``Elios 2 - indoor drone for confined space inspections.'' [Online]. Available:
  \url{http://www.flyability.com}
\BIBentrySTDinterwordspacing

\bibitem{23}
T.~Ozaslan, S.~J. Shen, Y.~Mulgaonkar, N.~Michael, and V.~Kumar, ``Inspection
  of penstocks and featureless tunnel-like environments using micro uavs,'' in
  \emph{Field and Service Robotics}, ser. Springer Tracts in Advanced Robotics,
  2015, Conference Proceedings.

\bibitem{24}
T.~{\"O}zaslan, K.~Mohta, J.~F. Keller, Y.~Mulgaonkar, C.~J. Taylor, V.~R.
  Kumar, J.~M. Wozencraft, and T.~Hood, ``Towards fully autonomous visual
  inspection of dark featureless dam penstocks using mavs,'' \emph{2016
  IEEE/RSJ International Conference on Intelligent Robots and Systems (IROS)},
  pp. 4998--5005, 2016.

\bibitem{25}
T.~{\"O}zaslan, G.~Loianno, J.~F. Keller, C.~J. Taylor, V.~R. Kumar, J.~M.
  Wozencraft, and T.~Hood, ``Autonomous navigation and mapping for inspection
  of penstocks and tunnels with mavs,'' \emph{IEEE Robotics and Automation
  Letters}, 2017.

\bibitem{26}
T.~{\"O}zaslan, G.~Loianno, J.~F. Keller, C.~J. Taylor, and V.~R. Kumar,
  ``Spatio-temporally smooth local mapping and state estimation inside
  generalized cylinders with micro aerial vehicles,'' \emph{IEEE Robotics and
  Automation Letters}, 2018.

\bibitem{28}
P.~Castillo, A.~Dzul, and R.~Lozano, ``Real-time stabilization and tracking of
  a four-rotor mini rotorcraft,'' \emph{IEEE Transactions on Control Systems
  Technology}, 2004.

\bibitem{27}
C.~H. Tan, D.~S.~B. Shaiful, W.~J. Ang, S.~K.~H. Win, and S.~Foong, ``Design
  optimization of sparse sensing array for extended aerial robot navigation in
  deep hazardous tunnels,'' \emph{IEEE Robotics and Automation Letters}, 2019.

\end{thebibliography}


\end{document}